\documentclass[a4paper,11pt]{article}
\usepackage{jinstpub} 
\usepackage{wasysym}
\usepackage{comment}
\usepackage{lineno}

\title{The DarkSide-20k experiment}

\author{A. Zani,}
\affiliation{INFN Milano, via Celoria 16, 20133 Milano, Italy}
\collaboration[c]{on behalf of the DarkSide-20k collaboration}

\emailAdd{andrea.zani@mi.infn.it}

\abstract{The DarkSide-20k experiment represents the present goal of the Global Argon Dark Matter Collaboration program. Bringing together the experience from previous argon-based detectors, as well as the knowledge gained on large volume membrane cryostats developed within the DUNE program, the community is now building a dual-phase LAr-TPC equipped with SiPM arrays for light readout. The main goal of the experiment is to discover or to extend the current sensitivity limits on the search for dark matter WIMP-like particles.
Currently, the experiment has entered the construction phase and the external cryostat is being put in place at Laboratori Nazionali del Gran Sasso (LNGS), Italy. Detector construction will follow, and data taking is expected to start in late 2026.
This contribution will introduce the DarkSide detector and goals, and it will report on the ongoing construction of the underground infrastructure at LNGS. Finally, it will concentrate on the current activities on large arrays of silicon light detectors, that are at the base of the construction of the detector light readout system.
}


\begin{document}
\maketitle
\flushbottom

\section{DarkSide-20k concept and physics goals}
\label{sec:intro}

The DarkSide~20k experiment (DS20k,~\cite{darkside20k}) is the current implementation of the Dark Matter (DM) search physics program carried out by the Global Argon Dark Matter Collaboration (GADMC). It aims at direct DM search in the form of Weakly Interactive Massive Particles, WIMPs. The physics case of the experiment is presented in~\cite{darkside20k}. The experiment exploits the technology of argon Dual-Phase TPC\footnote{Derived from the liquid argon TPC, LArTPC, concept~\cite{Rubbia:117852}.}, and it will be hosted deep underground at the Laboratori Nazionali del Gran Sasso (LNGS). It is an evolution of the DarkSide-50 detector~\cite{ds50} which has successfully operated at LNGS, obtaining the most sensitive results so far for an argon-based DM detector~\cite{ds50-dmsearch, ds-lowmass}.

Charged particles interacting in liquid argon (LAr) produce primary scintillation photons at 127~nm in vacuum (Vacuum Ultra Violet range, VUV,~\cite{Heindl_2010}) and ionization electrons. In a standard LArTPC the primary photons are detected (\textit{S1} signal), whereas the electrons are collected on an anode, by implementing a uniform and stable electric drift field between the anode and the cathode. In the dual-phase technology, the drift volume is vertical and topped by a pocket of argon gas (GAr); the electrons created in the liquid are extracted in the gas with a properly biased wire grid. The electric field in the gas region is set such that the extracted electrons produce an electroluminescence (light) signal, as they travel through the gas, before being collected on the anode. The primary and the secondary (\textit{S2}) light signals are collected by photo-detectors placed behind the transparent anode and cathode. 

Key to the sensitivity of a dark matter search is the avoidance and rejection of background. The shape of the \textit{S1} pulse, as well as the ratio \textit{S2}/\textit{S1}, are used to perform particle identification and distinguish potential DM candidate events (caused by nuclear recoils) from backgrounds such as heavily ionising particles (mainly alpha's) and electron recoils (electrons, gamma's)~\cite{ds50-dmsearch}.
Indeed, argon scintillation light has two components characterised by very different decay-time constants, at 6 and $\sim$1600~ns~\cite{DOKE199962}, therefore addressed in the community as \textit{fast} and \textit{slow}, respectively. The relative strength of these two components depends on the nature of the interacting particle: low-ionising particles produce light pulses that are dominated by the \textit{slow} component, whereas highly ionising events (among which, nuclear recoils) are characterised by \textit{fast} waveforms.
Some sources of background, however, cannot be distinguished from nuclear recoils via pulse shape discrimination. For example, neutron-induced events must be reduced with passive and active shielding: DS20k foresees two veto volumes (see section~\ref{sec:detector}), and it is shielded from cosmic rays interactions by sitting deep underground inside one of the caverns of the LNGS, at 3400~meter water equivalent (m.w.e.) depth. Coherent scattering from neutrinos instead represents an irreducible background.
Beyond cosmic and environmentat backgrounds, DM search experiments must minimize radiological backgrounds from construction materials: all the experiment components must be thoroughly selected via assaying for low radiopurity. Assembly must then proceed in a clean room environment with well defined handling and cleaning protocols.

\subsection{Atmospheric and Underground Argon}
\label{sec:argon}

One foundational strength of the DarkSide project (starting from DS10~\cite{ds10} and DS50~\cite{ds50}) is to further push the limit of radiopurity, by acting on the target material. One challenge for rare-event searches with argon-based detectors is the presence in atmospheric argon of $^{39}$Ar, a beta-decaying isotope with a spectrum endpoint at around 565~keV, half-life of 269~y and a concentration that produces about 1~Bq of decays per~kg. Its spectrum can significantly overlap with the Region-Of-Interest for DM searches. Furthermore, atmospheric argon shows presence of other radioactive isotopes produced by interactions with cosmic rays and human activities: $^{41}$Ar, $^{42}$Ar.

The DarkSide project has thus decided to exploit argon from underground (UAr): this should be protected from cosmic ray interactions and therefore much more radiopure than its atmospheric counterpart (AAr). Most underground argon, however, has similar amounts of $^{39}$Ar to atmospheric argon and only a long search allowed identifying a suitable well, in Colorado, US, from which to extract the UAr for DS20k~\cite{uar}.
Extraction will be performed with the dedicated URANIA plant, whose civil construction is starting now. It should become operative in late 2024, producing UAr at a rate of 250~kg/d, i.e., around 420~days will be needed to obtain the target mass for DS20k, that is 100~t plus some spare of liquid. 
Argon will be stored in liquid phase, then shipped to the Aria plant in Sardinia~\cite{aria}, where it will undergo isotopic distillation. From there, UAr will be shipped to LNGS for the filling of the experiment, but for small samples: these will instead go to Laboratorio Subterr\'aneo Canfranc (LSC), where the argon will be thoroughly characterized in the DArTinArDM detector~\cite{dart}.

\section{The DarkSide-20k detector}
\label{sec:detector}

The DS20k main detector is a LArTPC operated in dual-phase, with a vertical drift. The drift volume is defined within an octagonal barrel, with dimensions of about 348$h\, \times$ 350$\diameter$~cm. Electron extraction in gas is provided by a stainless steel (SS) grid made with $150\,\mu$m wires, with a pitch of 3~mm. The gas pocket thickness will be of 7~mm. The inner volume (drift volume + gas pocket) is enclosed in an acrylic octagon. The top and bottom lids of the octagon are made in pure acrylic and their inner planes are coated with a thin conductive material (Clevios$^{\mathrm{TM}}$) and a wavelength shifter, Tetra Phenyl Butadiene (TPB). The conductive layer allows biasing the inner planes of the two lids, such that they act as cathode (bottom one) and anode (top one).
TPB allows shifting the argon VUV scintillation light into the visible range (420~nm) for detection. The barrel pieces of the octagon are machined and coated with Clevios, to define a geometry of conductive rings (field cage) connected by resistors and biased such as to define a uniform and stable electric drift field of 200~V/cm. Slightly inward with respect to the barrel, thin acrylic layers supporting a dielectric mirror (ESR) coated with TPB ensure minimal light loss. \textit{S1} and \textit{S2} signals are collected by photo-detectors arranged in two Optical Planes above the anode and below the cathode of the TPC.

\begin{figure}[htbp]
\centering
\includegraphics[width=.48\textwidth]{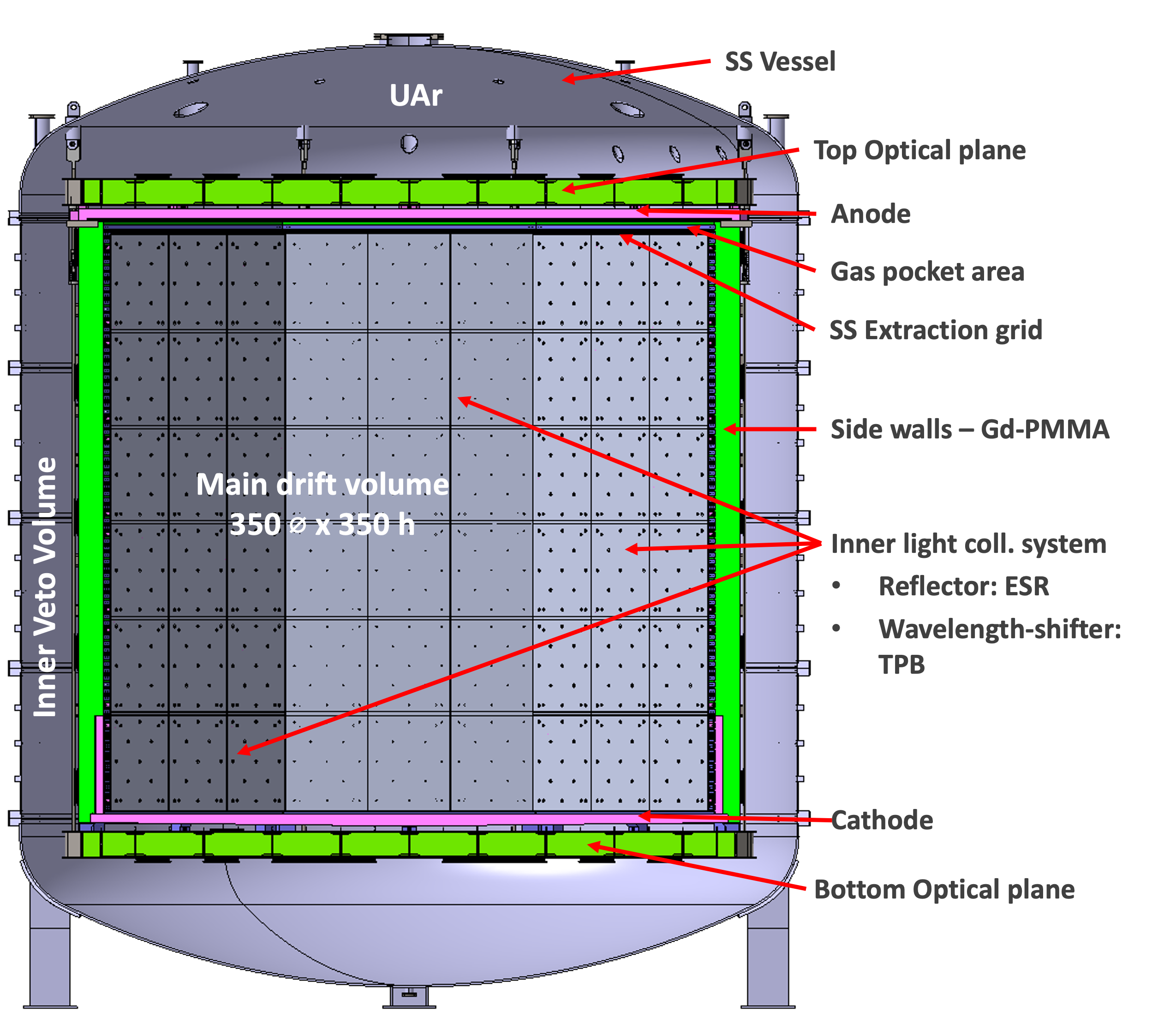}
\quad
\includegraphics[width=.48\textwidth]{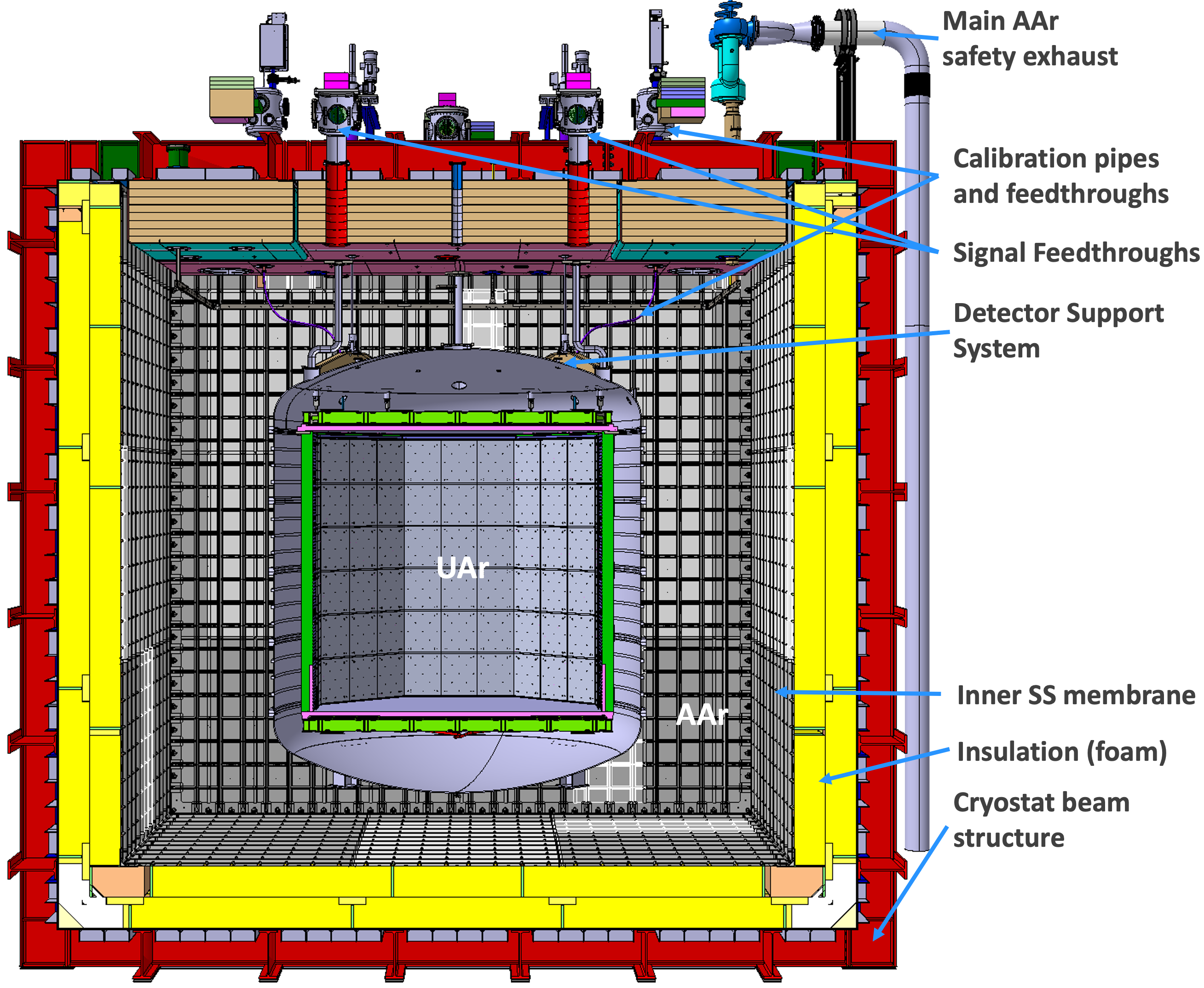}
\caption{Current design of the DarkSide-20k detector, shown in section views. Left, Inner Detector and Inner Veto, hosted in their SS vessel. Right, detector hanging on the roof of the AAr cryostat, along with installed feedthroughs for support and signal extraction. See text for details. \label{fig:design}}
\end{figure} 

This detector is inserted in a SS vessel filled with UAr (figure~\ref{fig:design}, left). The volume between the barrel and the vessel is sensitive as well, and it will act as Inner Veto: the TPC barrel panels are made with 15~cm thick acrylic doped with Gadolinium, uniformly dispersed within the plastic. Gd enhances neutron interactions, which then produce photons that can be detected in the TPC and in Inner Veto volume, with a different set of photo-detectors~\cite{gd-veto}. The back-planes of the Optical Planes also contain arrays of Gd-doped acrylic ``bricks'', for the same reason. With this design, the inner detector (inner veto) active UAr mass would be of 49.7~(32.0)~t, whereas the inner detector fiducial mass for the physics run would be 20.2~t.
The detector is equipped with a dedicated calibration system, made by a circuit of pipes running along the inner detector walls and flushed with nitrogen. A custom system is being designed to allow pushing/pulling radioactive sources through the pipes, to perform several measurements in different positions.

The SS vessel hosting the inner detector will be inserted in an AAr bath contained in a DUNE-like membrane cryostat~\cite{Montanari_2015}. A thin plastic passive shielding layer will be installed on the vessel outer surface, to moderate neutrons from the AAr cryostat and the LNGS cavern. The total AAr mass is of around 600~t and it will be used as an Outer Veto: 32~photo-detectors will be installed on the outer face of the plastic shielding, looking outwards, and they will be used to detect cosmic muons. The inner detector will be hanged through the SS vessel to the roof of the cryostat, via four supporting penetrations (figure~\ref{fig:design}, right).

\subsection{AAr cryostat and infrastructure construction}
\label{sec:cryostat}

\begin{figure}[htbp]
\centering
\includegraphics[width=.45\textwidth]{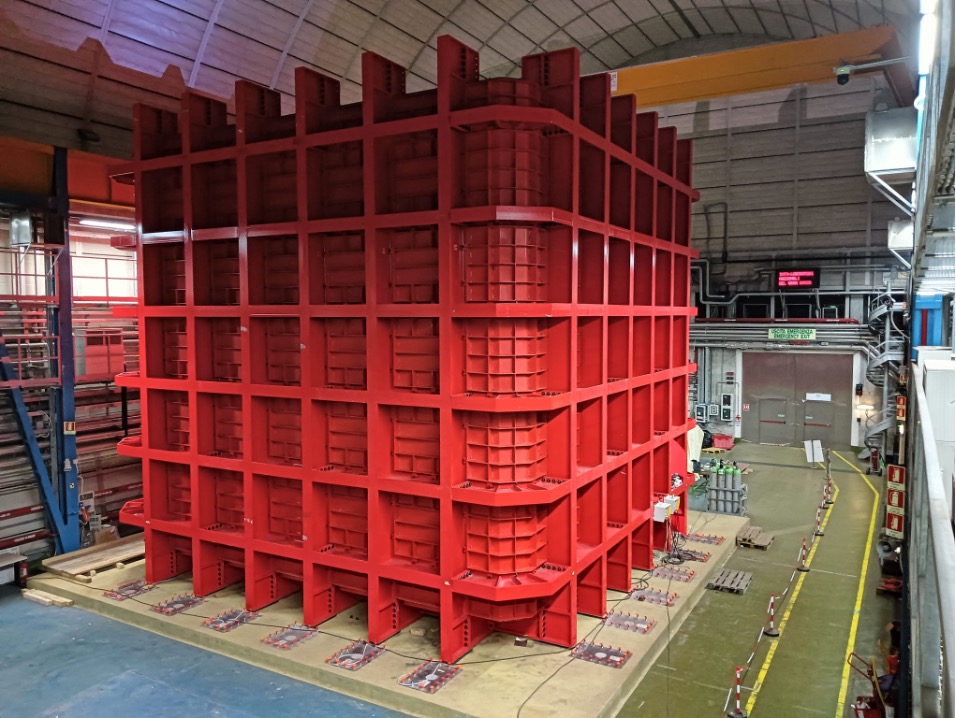}
\quad
\includegraphics[width=.5\textwidth]{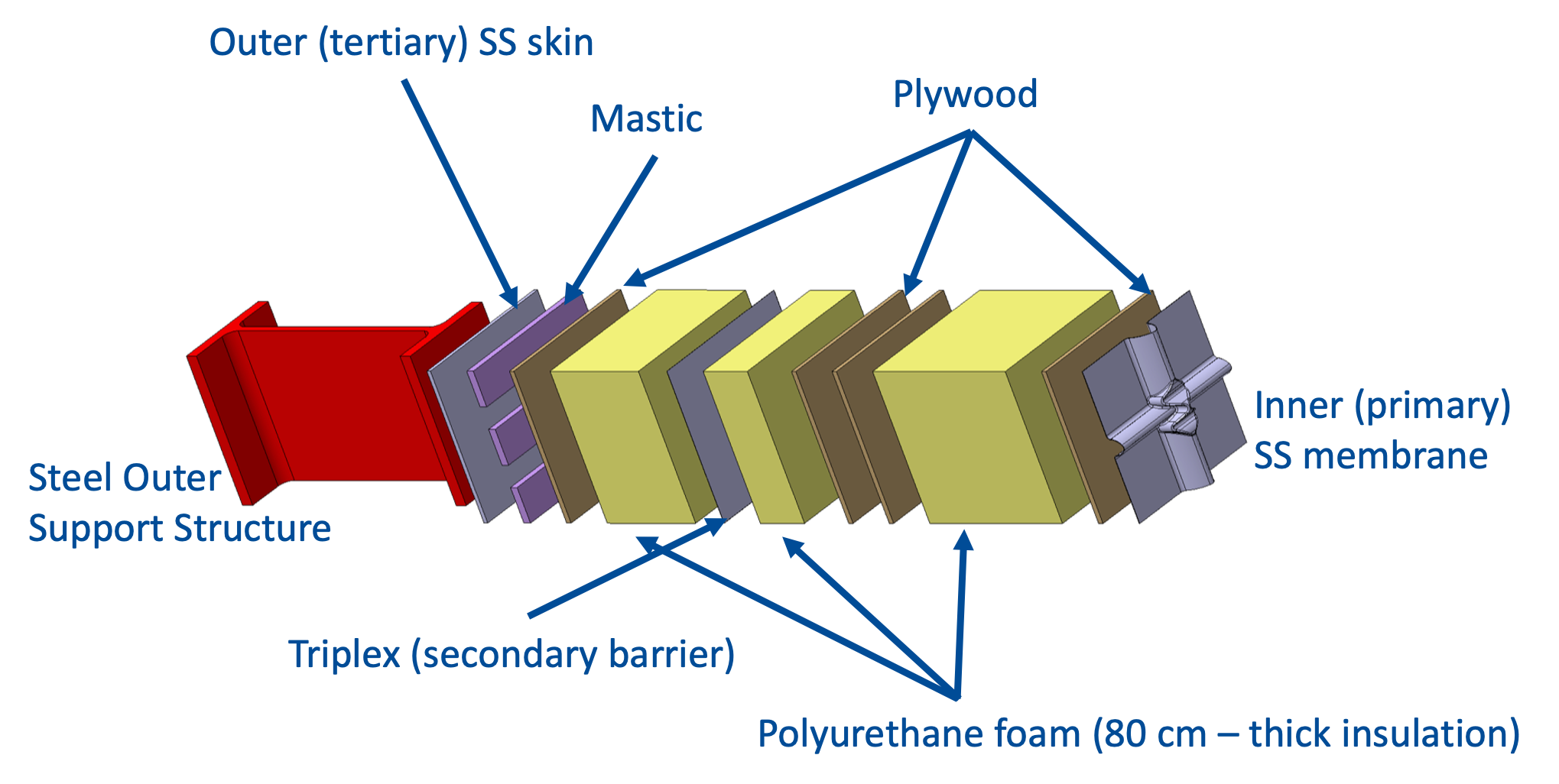}
\caption{Left, AAr cryostat external structure, completed, in LNGS Hall~C. Right, exploded view of the cryostat structure, including support beams, insulating foam and internal membrane. Details in text. \label{fig:cryostat}}
\end{figure}

The main AAr cryostat is currently under construction in the underground LNGS cavern called Hall~C. It is made with the same membrane technology already used for the ProtoDUNE detectors at CERN~\cite{pdsp_design, Abi_2020_PDSPperf, DuneDP_IDR, pddp_pds}, the SBND detector at FNAL~\cite{sbnd} and it is the technology of choice for DUNE~\cite{Abi_2020_V1}. As shown in figure~\ref{fig:cryostat}, the cryostat is made by an external support structure made of steel I-beams, enclosing an external SS ``tertiary'' membrane. Moving inside, two layers of polyurethane foam insulation are installed, separated by an intermediate ``secondary'' membrane, acting as barrier in case of AAr spill.
The innermost layer is the so-called ``primary'' membrane, a 1.2~mm thick sheet of corrugated SS: it acts as primary liquid containment barrier and its corrugations ensure high elasticity and capability for deformation during cool-down.
This cryostat technology is modular, thus allowing for exact prototyping (ProtoDUNE) and scalability. It is licensed by company GTT\footnote{\textit{Gaztransport \& Technigaz}, https://gtt.fr/ .},
and it is widely used for the sea transport of Liquid Natural Gas (LNG) on container ships.

The beam structure and the external membrane for DS20k were built in May-October~2023. Installation of insulation and inner membranes started in November~2023, until April~2024. The cryostat roof will be constructed outside LNGS, in 5 separate pieces ready to be ``plugged-in'' and complete with all the penetrations needed for the experiment. The cryostat is completed by a dedicated access structure, and by a cryogenic support structure, presently under construction and which will host the two cryogenic plants needed for the circulation and purification of UAr and AAr. In~2024, a custom clean room will be installed on top of the open cryostat, to allow for detector assembly inside; in 2024-2025 a three-storied control room will be constructed next to the cryostat.

\section{Photo-Detectors and NOA}
\label{sec:pe-noa}

\begin{figure}[htbp]
\centering
\includegraphics[width=.8\textwidth]{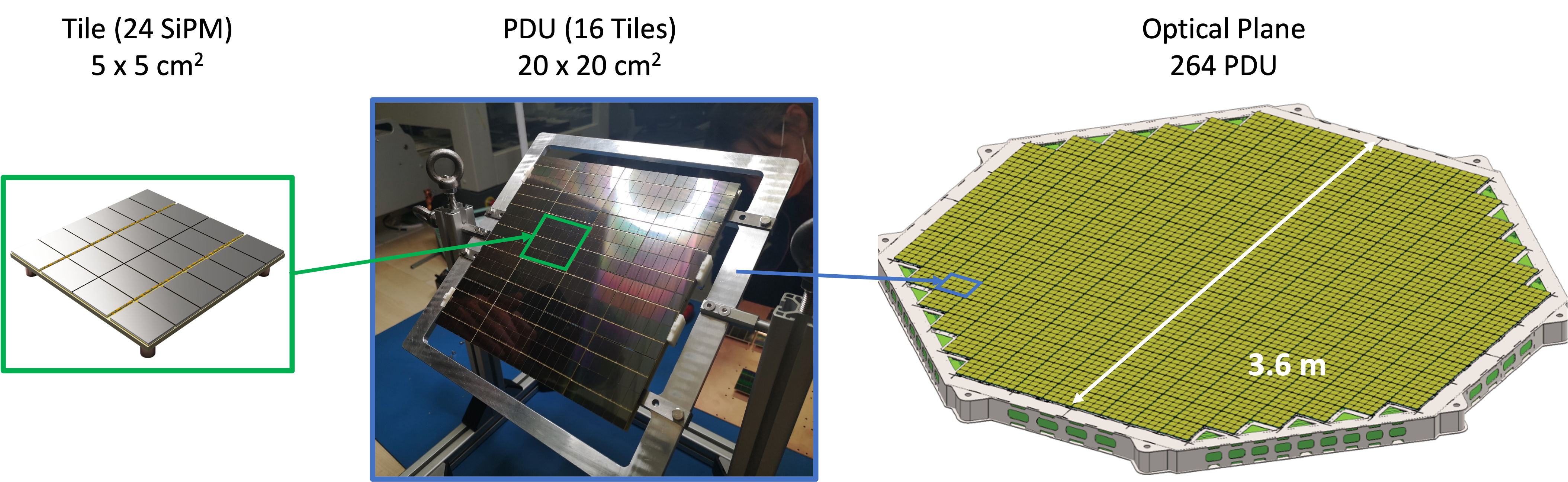}
\caption{Photon Detections Units in DS20k: 24~SiPMs are ganged together in a Tile (left); 16~Tiles are grouped in a PDU, i.e. four readout channels (centre); 264~PDUs are installed on each Optical Plane (right). \label{fig:pdu}}
\end{figure}

DS20k photon detection is based on Silicon PhotoMultipliers (SiPMs), in particular those developed with Fondazione Bruno Kessler\footnote{https://www.fbk.eu/en/ .}, Trento, IT. SiPMs can be preferred to PhotoMultiples Tubes (PMTs) for several reasons, among which lower bias voltage; insensitivity to magnetic fields; lower dark count rate (DCR) and high photon detection efficiency (PDE) at cryogenic temperature; and higher radio-purity. In particular, the devices selected for DS20k~\cite{sipm-charact} have PDE~$>40\%$ and DCR~$<0.01$~Hz/mm$^2$ at 77~K (at 7~V overVoltage). 

Single SiPMs ($8 \times 12$~mm$^2$) must be grouped together in order to obtain large-area photo-detectors. The basic light collection unit for DS20k is called Photon Detection Unit (PDU). To produce a PDU, one first has to gang together 24~SiPM in a $5 \times 5$~cm$^2$ ``Tile''; then, 16~Tiles are assembled into a $20 \times 20$~cm$^2$ PDU (see figure~\ref{fig:pdu}). Each unit presents four $10 \times 10$~cm$^2$ readout channels. 264~PDUs are installed in each Optical Plane, whereas 120~more (called ``veto PDUs'', vPDUs) are installed on the outer surfaces of the barrel pieces and of the Optical Planes (section~\ref{sec:detector}).

PDU are produced mostly in house by the Collaboration, starting from SiPM wafers from FBK/LFoundry. This is mainly done in a new INFN-LNGS facility, Nuova Officina Assergi (NOA) \cite{NOA-LC-TAUP2023}: this is a class ISO~6 clean room, made of two main rooms, dedicated to photo-detectors production and to large-volume detector assembly. The facility was commissioned in 2023, and it can be further upgraded by installing a dedicated Radon-abatement system.
Currently, NOA is populated with the machines needed to produce the DS20k SiPMs. 
Year 2023 was dedicated to the start-up of the activities for wafer characterization: 15\% of the received wafers was tested, resulting in a 90\% yield. 16~Tiles were produced and assembled into the first PDU, which was then tested in a dedicated test facility in Naples. TPC PDU production is planned to begin in May~2024, following completion of the pre-production, that started in late 2023.
Integration of the Optical Planes with the PDUs is then foreseen to happen in NOA in the first half of 2025.
Veto PDU production and testing is instead proceeding across collaborating institutions in UK and Poland~\cite{vpdu}.

\section{Outlook}
\label{sec:concl}

This contribution summarises the concept and the design of the DarkSide~20k experiment for Dark Matter direct detection. The detector technology and design builds upon the successful experience of its predecessors, as well as on the injection of new technologies, such as the membrane cryostat for the containment of the UAr in a thermal bath of AAr.
Construction of the experiment started in 2023, with the installation of the AAr cryostat in LNGS Hall~C, the commissioning of the NOA facility, and the start of SiPM production. At the same time, the contract for AAr cryogenics plant construction and installation was recently awarded, and there are ongoing tests in Hall~C on a mock-up set up, for the characterization of the performance of the UAr circulation plant. Overall, years 2023-2024 see multiple activities across all experiment components, with the final goal of completing the installation and being ready for commissioning at the end of~2026.

\bibliographystyle{JHEP}

\bibliography{biblio}

\end{document}